\documentclass[
twocolumn,
superscriptaddress,
 amsmath,amssymb,
 aps,floatfix
]{revtex4-1}
\usepackage{graphicx}
\usepackage{dcolumn}
\usepackage{bm}
\usepackage[USenglish]{babel}
\usepackage{appendix}
\usepackage{xcolor}
\definecolor{prlblue}{rgb}{0.176, 0.152, 0.57}
\usepackage[colorlinks=true, linkcolor=black, citecolor=prlblue, filecolor=black, urlcolor=prlblue]{hyperref}
\usepackage{cleveref}

\newcommand{\LOA}{LOA, ENSTA Paris, CNRS, Ecole Polytechnique, Institut Polytechnique de Paris, 91762 Palaiseau, France}
\newcommand{\Oslo}{Department of physics, University of Oslo, N-0316 Oslo, Norway}

\begin{document}
\title{High average gradient in a laser-gated multistage plasma wakefield accelerator}

\author{A.~Knetsch}
\email{alexander.knetsch@polytechnique.edu}
\affiliation{\LOA}

\author{I.A.~Andriyash}
\affiliation{\LOA}

\author{M.~Gilljohann}
\affiliation{\LOA}

\author{O. Kononenko}
\affiliation{\LOA}

\author{A.~Matheron}
\affiliation{\LOA}

\author{Y.~Mankovska}
\affiliation{\LOA}

\author{P. San Miguel Claveria}
\affiliation{\LOA}

\author{V.~Zakharova}
\affiliation{\LOA}

\author{E.~Adli}
\affiliation{\Oslo}

\author{S.~Corde}
\affiliation{\LOA}
\date{\today}

\begin{abstract}
Plasma wakefield accelerators driven by particle beams are capable of providing accelerating gradient several orders of magnitude higher than currently used radio-frequency technology, which could reduce the length of particle accelerators, with drastic influence on the development of future colliders at TeV energies and the minimization of x-ray free-electron lasers. Since inter-plasma components and distances are among the biggest contributors to the total accelerator length, the design of staged plasma accelerators is one of the most important outstanding questions in order to render this technology instrumental. Here, we present a novel concept to optimize inter-plasma distances in a staged beam-driven plasma accelerator by drive-beam coupling in the temporal domain and gating the accelerator via a femtosecond ionization laser. 
\end{abstract}

\maketitle
Electron particle accelerators regularly demonstrate great utility to a variety of scientific disciplines, such as in the form of x-ray free electron lasers for photon science and their applications or colliders for particle physics. Limited by the accelerating field of state-of-the-art accelerator modules, free-electron lasers driven by linear accelerators (linacs) already span several kilometers in length. To meet the requirements of future electron-positron colliders, the conventional linac-based machines with center-of-mass energies beyond 250 GeV and to several TeV, will reach tens of kilometers in length. 

Here, plasma wakefield accelerators (PWFAs) offer a promising alternative to the trend of ever-increasing particle accelerator sizes and towards compact X-ray-free electron lasers. 
Dense beams of relativistic charged particles can excite plasma wakes with accelerating gradients of 10-100 GV/m \cite{TajimaDawson1979}, in which a trailing electron beam can gain energy on much shorter distances. 
With such high fields inside the plasma accelerator, the resulting length of a linac becomes defined not as much by the size of the accelerator stage anymore, but rather by the general configuration of the machine. 
Therefore, a better figure of merit is the average accelerating gradient $E_\mathrm{avg}$, which is the ratio between the mean energy gained by a particle beam and the total length of the accelerator. 
For example, for the future international linear collider (ILC) based on conventional linac technology, this value estimates as $E_\mathrm{avg}^\mathrm{ILC} = 31.5\,\mathrm{MV}\mathrm{m}^{-1}$ for the overall length of 30-50~km\,\cite{ilc}. 
There is not a consensus yet, how large $E_\mathrm{avg}$ would be in the case of a PWFA collider facility, but the value should be close to 1 GV/m to be competitive.

In PWFAs, the energy from a drive beam is transferred to a trailing beam, such that an auxiliary accelerator as a source of the driver is required. If the trailing beam and drive beams are pre-accelerated to arrive at similar energies at a PWFA stage the trailing-beam energy can be doubled in a single plasma stage, such as demonstrated in the seminal work by Blumfeld et al., where the tail of a beam was accelerated in the wake, driven by its head\,\cite{Blumenfeld2007EnergyAccelerator}. 
These results motivate the concept of a PWFA afterburner, i.e. a plasma stage that can be added at the end of an accelerator as a final energy-doubling device.
However, even in such a desirable constellation,
the total length of the accelerator still remains tens-of-km-scale and the average gradient is only improved by a factor of approximately 2.
Using multiple stages, offers a more elegant solution\,\cite{adli2013design}. If a series of beams with much lower energy than the final trailing-beam energy is applied to transfer energy to the trailing beam in consecutive stages, the parts of the accelerator delivering the drive beams can be kept short and $E_\mathrm{avg}$ can be much higher\,\cite{adli2013design}.
Of course, this scheme comes with its own challenges. Inter-plasma components for in-coupling and out-coupling of drive-beams and transport of trailing beams between stages in most designs exceed the length of the plasma, strongly reduces $E_\mathrm{avg}$ and can be a source of emittance degradation.
Maintaining a high average accelerating gradient throughout the full length of an accelerator is therefore a major challenge of staging in PWFA when applied to TeV collider\,\cite{lindstrom2021staging}. 

 \begin{figure*}[ht]
    \includegraphics[width=0.9\textwidth]{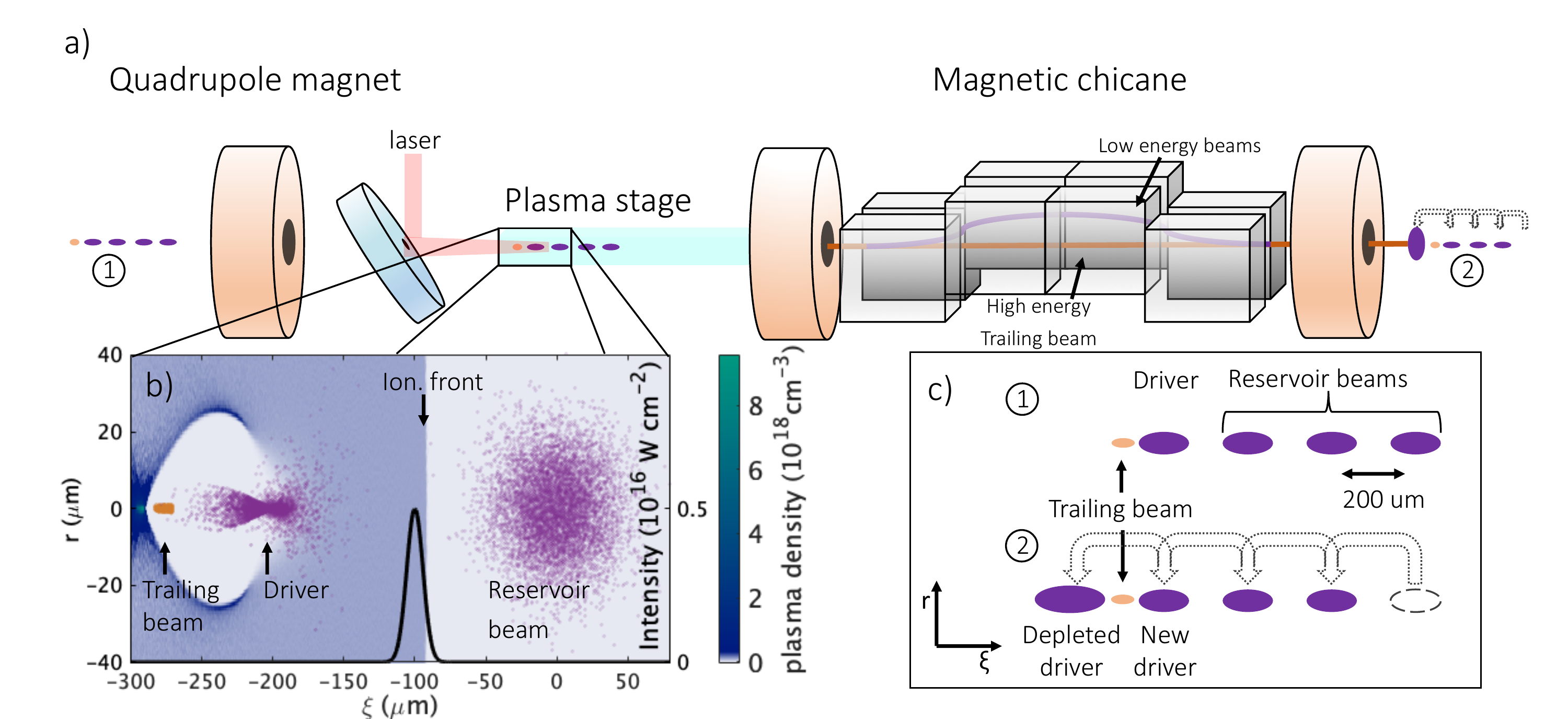}
    \caption{Acceleration stage (a) with of PWFA section, and chicane section to extract and insert new driver beams from beam reservoir. Driver and trailing beam participate in the PWFA process as they are later than the laser ionizer, while reservoir beams propagate through gas, as illustrated by a PIC-snapshot, showing reservoir beam, drive beam, trailing beam and a central cut through the plasma density. (b). The intensity of the fully resolved ionization laser is plotted in black. A sketch of the bunch-train configuration before and after a module is shown in (c).}
    \label{fig:concept}
\end{figure*}

Here, we suggest a staging method to reduce the footprint of inter-stage beamline-components and thereby presents a scalable concept with high average accelerating gradient. Instead of using parallel beamlines for drive and trailing beams, we propose to transport all beams on the same axis and perform in-coupling and out-coupling of drive beams in the temporal domain. 
This is possible because of a unique feature of laser-ionized PWFAs -- the ultra-short ionization front of femtosecond laser pulses.

The concept is illustrated in Figure\,\ref{fig:concept} and builds on three types of beams that propagate in one bunch train, but follow different orbits. The previously introduced \textit{driver} and \textit{trailing} beam, and beams that are not yet participating in a plasma accelerator, which we will call \textit{reservoir} beams in this context. After every plasma stage, a driver is used up and the trailing beam either gained energy or was refocused onto the next accelerating plasma stage in a beam-driven thin plasma lens\,\cite{DossPlasmaLensPRAB2019}. The reservoir 
beams form a bunch train with a small longitudinal spacing of $\Delta \xi_\mathrm{res}=200\,\mu$m in the co-moving coordinate $\xi=z-ct$,  ($c$ being the vacuum speed of light) and are transported by quadrupoles with alternating sign of magnetic field(a so-called focusing-defocusing or \textit{FODO} lattice). At the end of the reservoir-bunch train follows the drive beam and finally the trailing beam with a spacing of $\Delta \xi_\mathrm{tra}=75\,\mu\mathrm{m}$. Bunch trains with small spacing have been considered  to resonantly excite PWFAs\,\cite{manwani2021resonantBunchTrain} and even have been experimentally demonstrated before \cite{kallos2006resonantBunchTrain}.
As shown in a Particle-In-Cell (PIC) simulation snapshot in Figure\,\ref{fig:concept}b), the proposed spacing pattern is wide enough to fit a 23\,fs long laser pulse with a spot size of $w_0=100\,\mu\mathrm{m}$ and a peak intensity of $I_0=5\times 10^{15}\,\mathrm{W}\mathrm{cm}^{-2}$ between beams and thus gate-out reservoir beams from interacting with plasma via its short ionization front in a Hydrogen gas, while driver and trailing beam engage in a PWFA. To ionize a plasma channel over tens of centimeters, e.g. axicon lenses can be used. 
Every second FODO cell includes a weak chicane that delays reservoir beams by the length of their spacing and that way couples out the depleted driver to the back of the trailing beam.
Simultaneously, a fresh reservoir beam becomes a new driver for the subsequent plasma stage. An important advantage of this scheme, when compared to off-axis coupling strategies\,\cite{adli2013design} is that the trailing beam by design does not accumulate dispersion from the dipoles, which could otherwise increase its emittance.
It is also worth mentioning, that magnetic chicanes have been suggested before in the context of staged plasma accelerators with the aim to regulate energy spread of the trailing beams \cite{pousa2019compact,lindstrom2021self,pousa2022energy} and are used in laser-driven plasma wakefield accelerators driving free-electron lasers\,\cite{labat2022seeded,delbos2018lux}. In contrast to these concepts, the chicanes suggested here have a much weaker effect and are designed to leave the trailing beam unperturbed.

\begin{figure}[t]
    \includegraphics[width=0.51\textwidth]{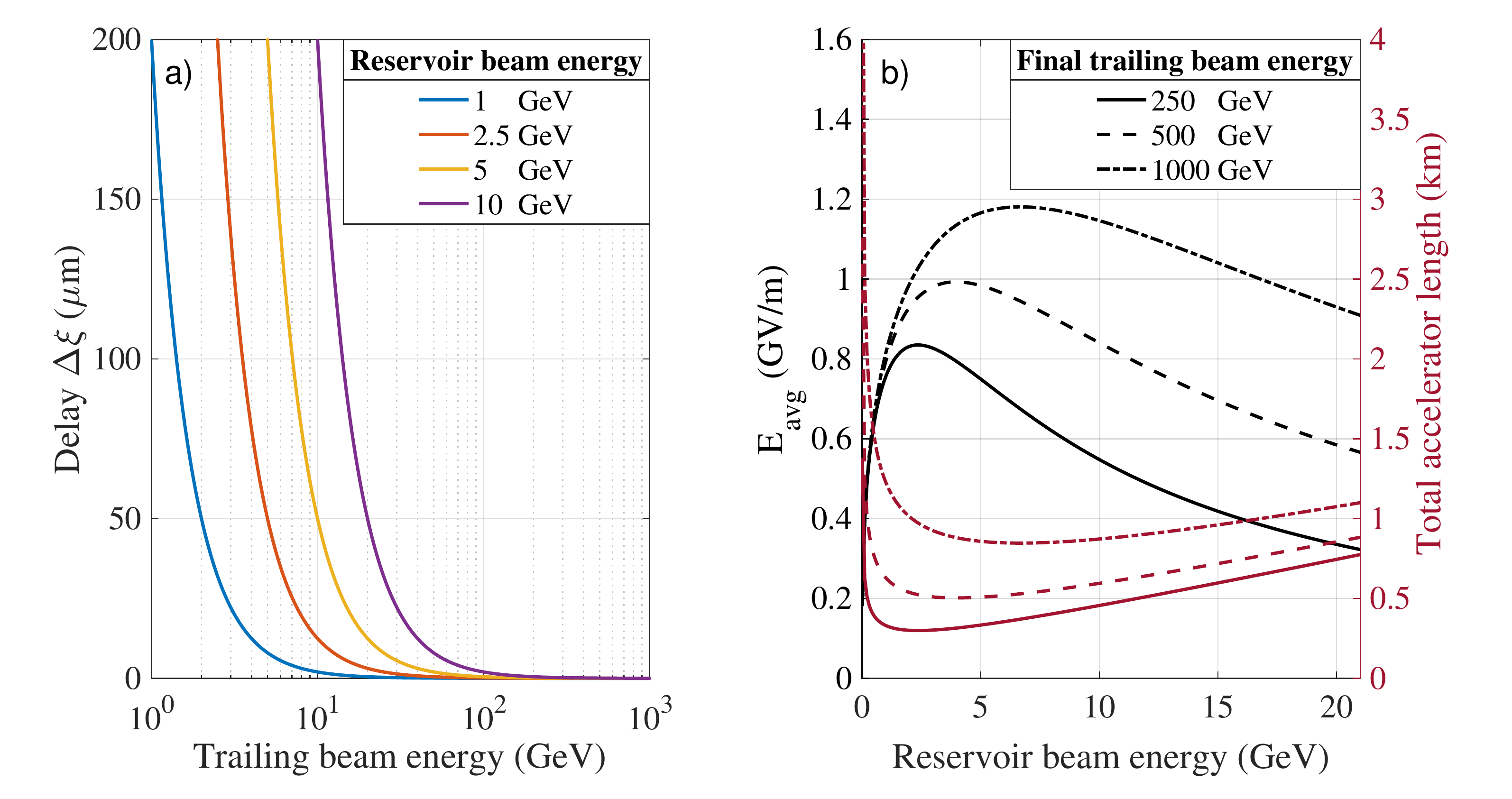}
    \caption{Trailing-beam delay accumulated in a chicane as a function of energy for different chicane designs, where the reservoir-beam delay is always kept at $200\,\mu$m (a). Total length (red) and corresponding average gradient (black) of a staged laser-gated PWFA accelerator, depending on reservoir-beam energy and final trailing-beam energy are plotted in (b).}
    \label{fig:Offset}
\end{figure}

To generate the desired delay, we propose to use a simple symmetric C-chicane consisting of four identical dipoles with magnetic field $B_\mathrm{D}$ and a dipole length of $L_\mathrm{D}$. 
We furthermore assume that the electron beams at play move at relativistic velocity, expressed by their Lorentz factor $\gamma$. 
Due to the longer path length of electrons in a chicane, the beams exit the chicane with a delay, here expressed as a length in the co-moving coordinate.
For a given delay, the overall chicane length is smallest for a minimum drift length between the dipoles, so we will set it to 0. In practice, there would of course be a small, but negligible gap between the magnets. Finally, the delay that a particle accumulates per chicane is 
\begin{equation}
    \Delta \xi=4L_\mathrm{D}\left(\sin^{-1}\left(\frac{L_\mathrm{D}}{R_\mathrm{gyr.}}\right)\frac{R_\mathrm{gyr.}}{L_\mathrm{D}}-1\right),
    \label{eq:Offset}
\end{equation}
where $R_\mathrm{gyr.}= \frac{\gamma m_\mathrm{e}c}{q_\mathrm{e}B_\mathrm{D}}$ is the gyration radius for electrons with mass $m_\mathrm{e}$ and charge $q_\mathrm{e}$ inside the magnetic field.
We can see that it is important that the trailing beam is at higher energy than the reservoir beams to remain unaffected by the chicanes. This requirement is explored quantitatively in Figure\,\ref{fig:Offset}\,a), where Eq.\,(\ref{eq:Offset}) is evaluated for different chicane parameters and trailing-beam energies. For example, in chicanes that delay reservoir beams with an energy as low as 1 GeV by $200\,\mu\mathrm{m}$, a trailing beam at an energy higher than 14 GeV accumulates a delay of less than $1\,\mu\mathrm{m}$. Considering a plasma density of $10^{17}\,\mathrm{cm}^{-3}$ with a corresponding plasma wavelength of $106\,\mu\mathrm{m}$ this is equivalent to a phase-offset of $\approx 0.5^{\circ}$ and can easily be compensated by fine-adjusting the chicane parameters. In the important energy range of trailing-beam energies between 0.1 TeV and 1 TeV, trailing-beam delays in $\xi$ converge towards 0 such that chicanes that are located further downstream in the linac can effectively be built self-similar as soon as the trailing beam-energy approaches a ten-fold of the reservoir-beam energy. 
As a next step, reservoir-beam energies are optimized to maximize $E_\mathrm{avg}$.
Reservoir beams at relatively low energies can be provided by a small pre-accelerator, but such a design requires many stages, which increases complexity and distance between stages and ultimately decreases the average gradient. On the other extreme, large reservoir-beam energies can accelerate beams in only a few stages, but need longer chicanes and pre-accelerators that contribute to an overall large accelerator footprint.
To find the optimum conditions, we calculate the total length of the accelerator,
\begin{equation}
\label{eqn:acclength}
    L_\mathrm{acc.}=\frac{4 L_\mathrm{chicane}W_\mathrm{final}}{T W_\mathrm{driver}}+\frac{W_\mathrm{driver}}{E_\mathrm{avg.,pre}}.
\end{equation}  
The length of one repeating stage is at least four times the chicane length $L_\mathrm{chicane}$. Then, the number of stages required to reach the desired final trailing-beam energy $W_\mathrm{final}$ depends on the fraction of the drive-beam energy $W_\mathrm{driver}$
that the trailing beam gains in one PWFA stage $T=\frac{\Delta W_\mathrm{tra.}}{W_\mathrm{driver}}$. We assume $T\overset{!}{=}2$ as justified by PIC simulations.
For simplification purpose, the length does not consider a final focusing of a potential collider or undulator sections in case of an FEL.
The second term describes the length of the pre-accelerator. Here, we use the average gradient planned for the ILC, but in the future, $E_\mathrm{avg.,pre}$ could be increased e.g. by applying X-band cavities\,\cite{cahill2018high}. 

Equation (\ref{eqn:acclength}) is plotted in Fig.\,\ref{fig:Offset} b) and demonstrates that optimal conditions vary, depending on the target energy. 
The calculations for Fig.\,\ref{fig:Offset} b) are performed for dipoles with a magnetic field of 1\,T. 
It is worth mentioning, that permanent dipoles with magnetic strengths $> 2\,\mathrm{T}$ have been built before by employing a Halbach-like configuration\,\cite{kumada2003strongest}, so a chicane with larger magnetic fields can be built without the need for superconducting components. With a
higher magnetic fields the chicanes could be built smaller and the average gradient would be further increased. The results based on simple scaling-laws indicate that an accelerator consisting of laser-gated PWFAs could exceed an average gradient of 1 GV/m.

\begin{figure}[b]
    \includegraphics[width=0.5\textwidth]{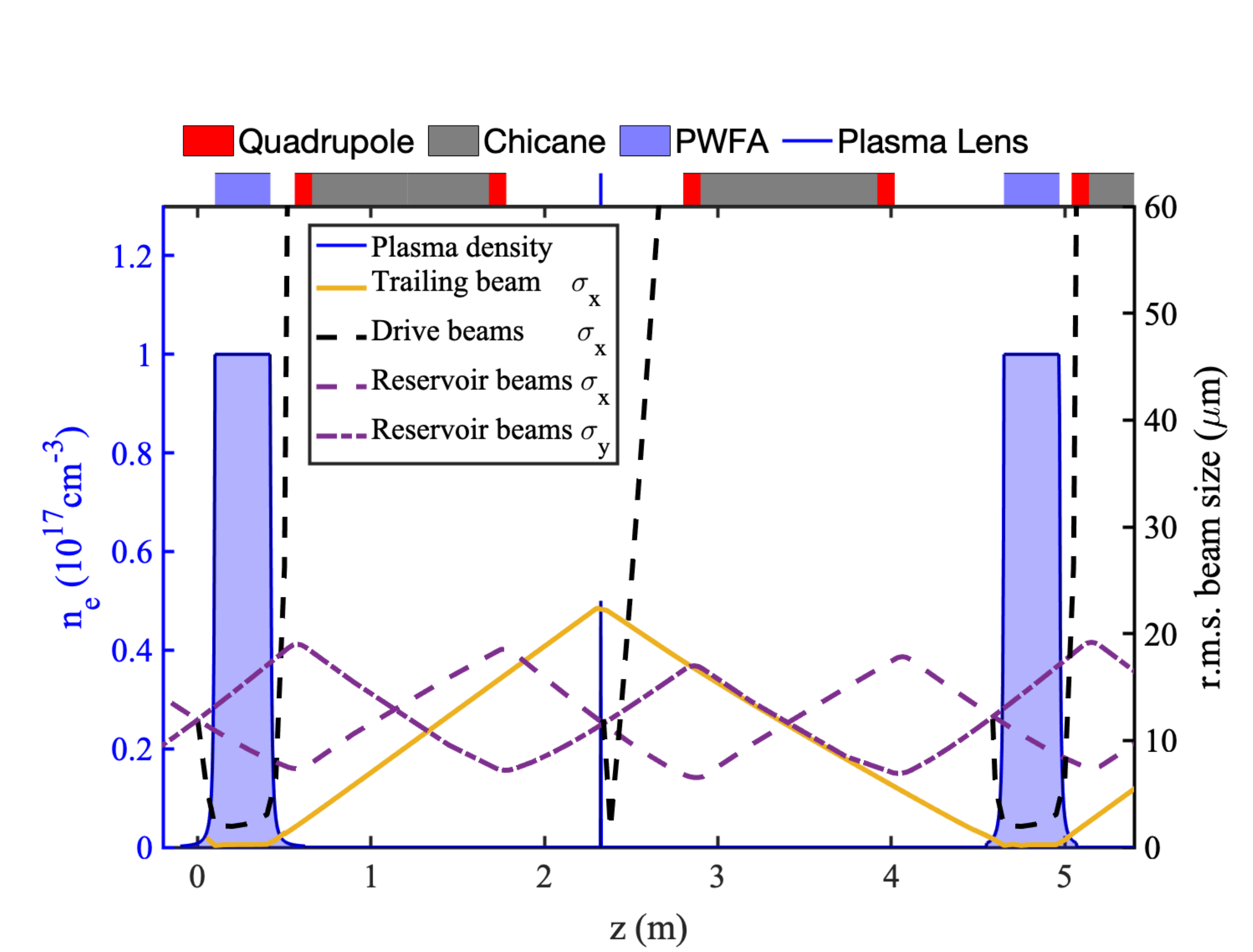}
    \caption{Transport of different beam types between plasma stages. The reservoir beams propagate in a FODO lattice and are not influenced by plasma wakes, as they arrive too early. Drive beams lose energy to the plasma wake and leave the plasma with increased divergence. The trailing beam is refocused by a plasma lens approximately centered between the PWFA stages.}
    \label{fig:ReservoirLattice}
\end{figure}

To demonstrate the proposed concept, a system of two consecutive plasma accelerator stages and one plasma lens stage was modeled numerically from start to end.
Figure \ref{fig:ReservoirLattice} illustrates the transport of the different beam types throughout this system. 
Reservoir beams with an energy of $2.5\,\mathrm{GeV}$ and an energy spread of $0.8\%$ full width at half maximum (FWHM) follow the magnetic lattice, i.e. set by dipoles and quadrupole magnets with a magnetic field strength of 112 T/m and a length of 10 cm. The quadrupole magnets are placed at a distance of 1.12 m from one another, and dipoles which fill every second FODO half-cell have a length of 0.275 m and a magnetic field of $1\,$T each. Simulations for the reservoir-beam propagation were performed with the code elegant\,\cite{borland2000elegant} and optimized such that a drive beam is radially symmetric at the entrance of a plasma with a spot size of $12\,\mu\mathrm{m}$. 

\begin{figure}[ht]
    \includegraphics[width=0.5\textwidth]{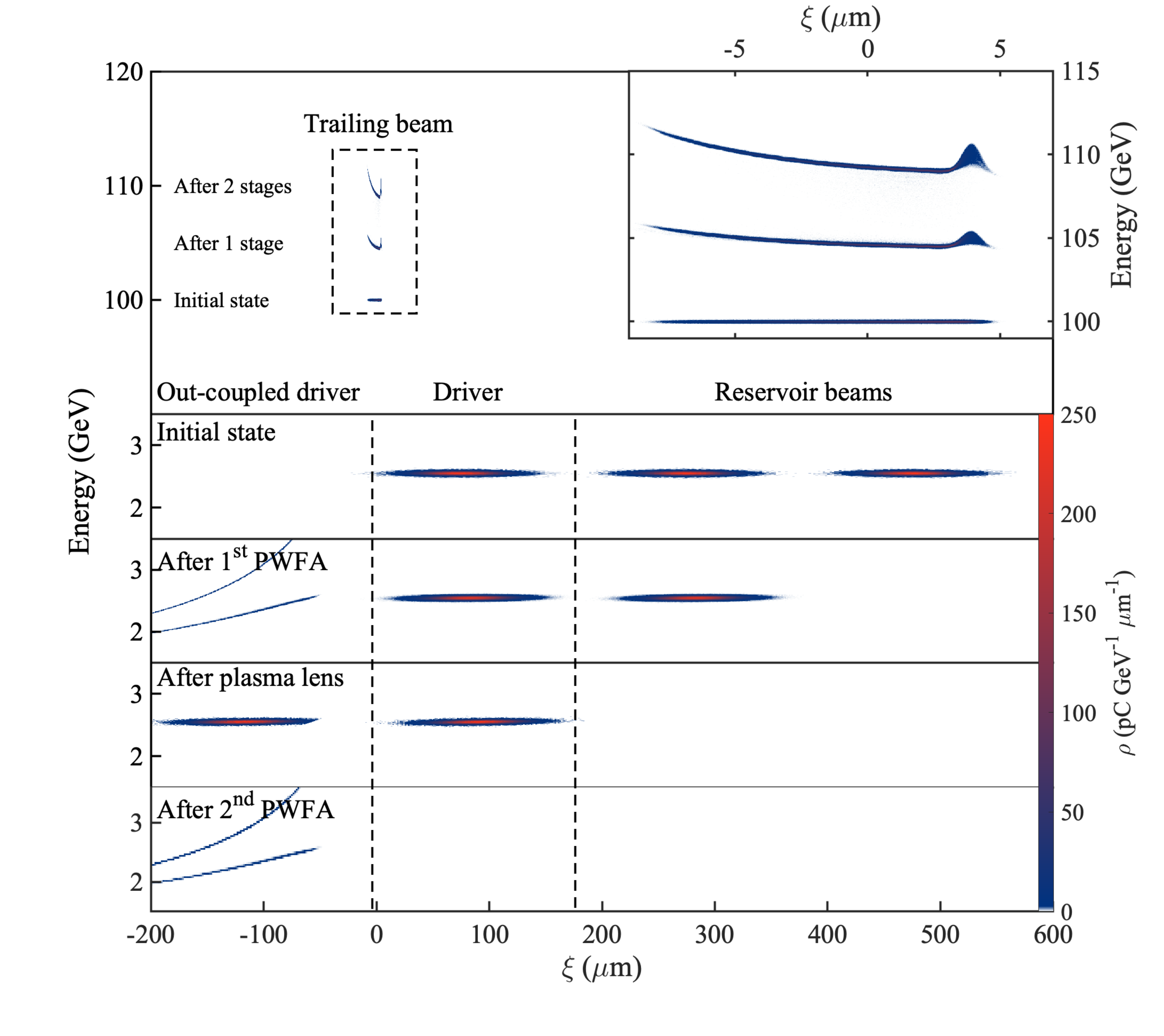}
    \caption{Numerical modeling of the longitudinal phase space at different locations. Trailing beam distributions are plotted as initial state and after accelerating sections (top) and as zoom (inset). Modeled macro-particle distributions of driver and reservoir beams and macro-particle output from PIC are plotted. All macro-particle distributions are delayed following Eq.\,(\ref{eq:Offset}).}
    \label{fig:Sims_electrons}
\end{figure}

The plasma stages are simulated with the quasi-3D PIC code FBPIC\,\cite{Lehe2016FBPIC}. Between the plasma stages, PIC-simulated macro particles of trailing and drive beams are transported according to linear beam-propagation matrices and delayed following Eq.\,(\ref{eq:Offset}). The FBPIC simulations were run in a boosted frame with Lorentz factor of 1.5 and a resolution of $\Delta r = 0.25\,\mu\mathrm{m},\Delta z = 0.225\,\mu\mathrm{m}$ with one azimuthal mode in a box size of $180\,\mu\mathrm{m} \times 75\,\mu\mathrm{m}$.
The simulated 2.5\,GeV reservoir beams have a charge of 700 pC and a root-mean-square (r.m.s.) bunch length of 27\,fs  to excite a plasma wake at the plateau plasma density of 
$1\times10^{17}\,\mathrm{cm}^{-3}$.  As confirmed by PIC simulations, such beams are not dense enough to tunnel-ionize hydrogen to plasma\,\cite{ADK_original}. 
The trailing beam with a charge of 80\,pC and an initial energy of 100\,GeV has a total bunch length of 12\,$\mu \mathrm{m}$ and a triangular current shape to decrease the growth of energy spread of $0.2\%$. 
Matching of the trailing beam to the transverse plasma wakefields, is facilitated by long plasma density ramps at entrance and exit of the accelerating stages, but ultimately requires a refocusing element between the accelerating PWFA stages. 
The plasma density of the density ramps follows the function, $n_e(z)=\frac{1}{(1-z/L_\mathrm{r})^2}$ as suggested by Xu et al.\,\cite{xu2016AdiabaticRamps} with a length scale $L_\mathrm{r}=0.1\,\mathrm{m}$ and a flat-top length of 31 cm. 
The positive effect of long density ramps at the end of capillary gas targets of tens of cm have been experimentally observed recently, with promising guiding effect on trailing beams \cite{knetsch2021stable}, however, long ramps reduce the overall energy gain, because the beam slips into the decelerating phase of the wake at low densities. 
To transport the trailing beam from one PWFA stage to the next, a beam-driven plasma lens is considered as suggested in \cite{DossPlasmaLensPRAB2019}. At the cost of a drive-beam for every plasma lens, this concept allows for strong linear focusing fields, which is advantageous to avoid emittance growth.
The plasma lens in this design with a length of $L_\mathrm{lens}=800\,\mu\mathrm{m}$ and a peak plasma density $1\times 10^{16}\,\mathrm{cm}^{-3}$
images the beam from the exit of one PWFA to the entrance of the subsequent one with a magnification close to 1 and matching is facilitated by the upramp.
The simulated focal length roughly follows the theoretical scaling law $f=\frac{2\gamma}{k_\mathrm{p}^2 L_\mathrm{lens}}\propto \frac{\gamma}{L_\mathrm{lens}n_\mathrm{e}}$.
It is straightforward to see that $f$ can be kept constant up to a trailing-beam energy of 1 TeV, either by scaling up the plasma lens density or by increasing $L_\mathrm{lens}$. Since $L_\mathrm{lens}$ also changes the distance between both foci, it is possible to fine-tune magnification between the stages and focal length independently. 
Figure \ref{fig:Sims_electrons} explores the development of the longitudinal phase space of different beam types throughout two PWFA stages and a plasma lens.
After one PWFA stage, the drive beam loses the majority of its energy, with a mean energy loss of 1.8 GeV. 
All drive-beam electrons (in a PWFA or plasma lens) are delayed to  $\xi<0$ and are, as far as the trailing beam is concerned, out-coupled. Due to its wide divergence of 5.7 mrad, the drive-beam charge expands radially and would, in an actual machine, be dumped into absorbing material located around the beam pipe. 
Simultaneously, the driver is replenished by delaying the reservoir beams, which need to be at per-mille energy spread to avoid smearing out the bunch train throughout many stages.
The longitudinal phase space of the trailing beam is plotted in its initial state, after one accelerating PWFA stage and after the second. The plasma lens is too short to influence the beam energy significantly and its effect is therefore not plotted.
Over the path of two PWFA stages, the trailing beam gains 9.28 GeV,
which puts $T\approx1.9$, close to the previously assumed value of 2.  
The achieved values set the average gradient of two accelerator stages to $1.04\,\mathrm{GV/m}$.
When considering a pre-accelerator with an average field of $E_\mathrm{avg}^\mathrm{ILC}=31.5\,\mathrm{MV m}^{-1}$, a laser-gated PWFA accelerator would span approximately 1050 m and be driven by a 288 ps long reservoir-beam consisting of 432 bunches. Such a machine would be capable to provide electron beams with an energy of 1 TeV at a total average accelerating gradient of $0.96\,\mathrm{GV/m}$.

In conclusion, the proposed concept of a laser-gated multistage beam-driven plasma accelerator addresses outstanding questions on how a plasma wakefield accelerator can be scaled up to the TeV energy range at a competitive average gradient. 
By in- and out-coupling the drive beam in the temporal domain and gating the accelerator with fs ionization lasers, this strategy provides a unique opportunity to combine plasma-based and magnet-based lattices on the same axis and thereby reduce the overall accelerator footprint to a few-kilometer scale. 
Presented results open the way towards compact gamma-gamma\,\cite{telnov1995principles} and electron-electron\,\cite{yakimenko2019prospect} colliders, and, by being compatible with laser-ionized beam-driven positron accelerators\,\cite{diederichs2019positron}, to $e^+$ $e^-$ colliders, that are strongly desired for future discoveries in the field of particle and high energy physics.

\begin{acknowledgments}
We acknowledge the Grand Équipement National de Calcul Intensif (GENCI) Joliot-Curie et Très Grand Centre de Calcul (TGCC) for granting us access to the supercomputer IRENE under Grants No. 2020-A0080510786, 2021-A0100510786, 2021-A0110510062 and 2022-A0120510786 to run PIC simulations. The work was supported by the European Research Council (ERC) under the European Union’s Horizon 2020 research and innovation programme (M-PAC project, Grant Agreement No. 715807).
\end{acknowledgments}


\begin{thebibliography}{23}%
\makeatletter
\providecommand \@ifxundefined [1]{%
 \@ifx{#1\undefined}
}%
\providecommand \@ifnum [1]{%
 \ifnum #1\expandafter \@firstoftwo
 \else \expandafter \@secondoftwo
 \fi
}%
\providecommand \@ifx [1]{%
 \ifx #1\expandafter \@firstoftwo
 \else \expandafter \@secondoftwo
 \fi
}%
\providecommand \natexlab [1]{#1}%
\providecommand \enquote  [1]{``#1''}%
\providecommand \bibnamefont  [1]{#1}%
\providecommand \bibfnamefont [1]{#1}%
\providecommand \citenamefont [1]{#1}%
\providecommand \href@noop [0]{\@secondoftwo}%
\providecommand \href [0]{\begingroup \@sanitize@url \@href}%
\providecommand \@href[1]{\@@startlink{#1}\@@href}%
\providecommand \@@href[1]{\endgroup#1\@@endlink}%
\providecommand \@sanitize@url [0]{\catcode `\\12\catcode `\$12\catcode
  `\&12\catcode `\#12\catcode `\^12\catcode `\_12\catcode `\%12\relax}%
\providecommand \@@startlink[1]{}%
\providecommand \@@endlink[0]{}%
\providecommand \url  [0]{\begingroup\@sanitize@url \@url }%
\providecommand \@url [1]{\endgroup\@href {#1}{\urlprefix }}%
\providecommand \urlprefix  [0]{URL }%
\providecommand \Eprint [0]{\href }%
\providecommand \doibase [0]{http://dx.doi.org/}%
\providecommand \selectlanguage [0]{\@gobble}%
\providecommand \bibinfo  [0]{\@secondoftwo}%
\providecommand \bibfield  [0]{\@secondoftwo}%
\providecommand \translation [1]{[#1]}%
\providecommand \BibitemOpen [0]{}%
\providecommand \bibitemStop [0]{}%
\providecommand \bibitemNoStop [0]{.\EOS\space}%
\providecommand \EOS [0]{\spacefactor3000\relax}%
\providecommand \BibitemShut  [1]{\csname bibitem#1\endcsname}%
\let\auto@bib@innerbib\@empty
\bibitem [{\citenamefont {Tajima}\ and\ \citenamefont
  {Dawson}(1979)}]{TajimaDawson1979}%
  \BibitemOpen
  \bibfield  {author} {\bibinfo {author} {\bibfnamefont {T.}~\bibnamefont
  {Tajima}}\ and\ \bibinfo {author} {\bibfnamefont {J.~M.}\ \bibnamefont
  {Dawson}},\ }\href {\doibase 10.1103/PhysRevLett.43.267} {\bibfield
  {journal} {\bibinfo  {journal} {Phys. Rev. Lett.}\ }\textbf {\bibinfo
  {volume} {43}},\ \bibinfo {pages} {267} (\bibinfo {year} {1979})}\BibitemShut
  {NoStop}%
\bibitem [{\citenamefont {Behnke}\ \emph {et~al.}(2013)\citenamefont {Behnke},
  \citenamefont {Brau}, \citenamefont {Foster}, \citenamefont {Fuster},
  \citenamefont {Harrison}, \citenamefont {Paterson}, \citenamefont {Peskin},
  \citenamefont {Stanitzki}, \citenamefont {Walker},\ and\ \citenamefont
  {Yamamoto}}]{ilc}%
  \BibitemOpen
  \bibinfo {editor} {\bibfnamefont {T.}~\bibnamefont {Behnke}}, \bibinfo
  {editor} {\bibfnamefont {J.~E.}\ \bibnamefont {Brau}}, \bibinfo {editor}
  {\bibfnamefont {B.}~\bibnamefont {Foster}}, \bibinfo {editor} {\bibfnamefont
  {J.}~\bibnamefont {Fuster}}, \bibinfo {editor} {\bibfnamefont
  {M.}~\bibnamefont {Harrison}}, \bibinfo {editor} {\bibfnamefont {J.~M.}\
  \bibnamefont {Paterson}}, \bibinfo {editor} {\bibfnamefont {M.}~\bibnamefont
  {Peskin}}, \bibinfo {editor} {\bibfnamefont {M.}~\bibnamefont {Stanitzki}},
  \bibinfo {editor} {\bibfnamefont {N.}~\bibnamefont {Walker}}, \ and\ \bibinfo
  {editor} {\bibfnamefont {H.}~\bibnamefont {Yamamoto}},\ eds.,\ \href
  {https://arxiv.org/abs/1306.6327} {\emph {\bibinfo {title} {The International
  Linear Collider}}},\ \bibinfo {number} {Technical Design Report}\ (\bibinfo
  {year} {2013})\BibitemShut {NoStop}%
\bibitem [{\citenamefont {Blumenfeld}\ \emph {et~al.}(2007)\citenamefont
  {Blumenfeld}, \citenamefont {Clayton}, \citenamefont {Decker}, \citenamefont
  {Hogan}, \citenamefont {Huang}, \citenamefont {Ischebeck}, \citenamefont
  {Iverson}, \citenamefont {Joshi}, \citenamefont {Katsouleas}, \citenamefont
  {Kirby} \emph {et~al.}}]{Blumenfeld2007EnergyAccelerator}%
  \BibitemOpen
  \bibfield  {author} {\bibinfo {author} {\bibfnamefont {I.}~\bibnamefont
  {Blumenfeld}}, \bibinfo {author} {\bibfnamefont {C.~E.}\ \bibnamefont
  {Clayton}}, \bibinfo {author} {\bibfnamefont {F.~J.}\ \bibnamefont {Decker}},
  \bibinfo {author} {\bibfnamefont {M.~J.}\ \bibnamefont {Hogan}}, \bibinfo
  {author} {\bibfnamefont {C.}~\bibnamefont {Huang}}, \bibinfo {author}
  {\bibfnamefont {R.}~\bibnamefont {Ischebeck}}, \bibinfo {author}
  {\bibfnamefont {R.}~\bibnamefont {Iverson}}, \bibinfo {author} {\bibfnamefont
  {C.}~\bibnamefont {Joshi}}, \bibinfo {author} {\bibfnamefont
  {T.}~\bibnamefont {Katsouleas}}, \bibinfo {author} {\bibfnamefont
  {N.}~\bibnamefont {Kirby}},  \emph {et~al.},\ }\href {\doibase
  10.1038/nature05538} {\bibfield  {journal} {\bibinfo  {journal} {Nature}\
  }\textbf {\bibinfo {volume} {445}},\ \bibinfo {pages} {741} (\bibinfo {year}
  {2007})}\BibitemShut {NoStop}%
\bibitem [{\citenamefont {Adli}\ \emph {et~al.}(2013)\citenamefont {Adli},
  \citenamefont {Delahaye}, \citenamefont {Gessner}, \citenamefont {Hogan},
  \citenamefont {Raubenheimer}, \citenamefont {An}, \citenamefont {Mori},\ and\
  \citenamefont {Joshi}}]{adli2013design}%
  \BibitemOpen
  \bibfield  {author} {\bibinfo {author} {\bibfnamefont {E.}~\bibnamefont
  {Adli}}, \bibinfo {author} {\bibfnamefont {J.}~\bibnamefont {Delahaye}},
  \bibinfo {author} {\bibfnamefont {S.}~\bibnamefont {Gessner}}, \bibinfo
  {author} {\bibfnamefont {M.}~\bibnamefont {Hogan}}, \bibinfo {author}
  {\bibfnamefont {T.}~\bibnamefont {Raubenheimer}}, \bibinfo {author}
  {\bibfnamefont {W.}~\bibnamefont {An}}, \bibinfo {author} {\bibfnamefont
  {W.}~\bibnamefont {Mori}}, \ and\ \bibinfo {author} {\bibfnamefont
  {C.}~\bibnamefont {Joshi}},\ }\href
  {https://accelconf.web.cern.ch/IPAC2013/papers/tupme020.pdf} {\bibfield
  {journal} {\bibinfo  {journal} {Proc. IPAC13}\ } (\bibinfo {year}
  {2013})}\BibitemShut {NoStop}%
\bibitem [{\citenamefont
  {Lindstr{\o}m}(2021{\natexlab{a}})}]{lindstrom2021staging}%
  \BibitemOpen
  \bibfield  {author} {\bibinfo {author} {\bibfnamefont {C.~A.}\ \bibnamefont
  {Lindstr{\o}m}},\ }\href
  {https://journals.aps.org/prab/abstract/10.1103/PhysRevAccelBeams.24.014801}
  {\bibfield  {journal} {\bibinfo  {journal} {Phys. Rev. Accel. Beams}\
  }\textbf {\bibinfo {volume} {24}},\ \bibinfo {pages} {014801} (\bibinfo
  {year} {2021}{\natexlab{a}})}\BibitemShut {NoStop}%
\bibitem [{\citenamefont {Doss}\ \emph {et~al.}(2019)\citenamefont {Doss},
  \citenamefont {Adli}, \citenamefont {Ariniello}, \citenamefont {Cary},
  \citenamefont {Corde}, \citenamefont {Hidding} \emph
  {et~al.}}]{DossPlasmaLensPRAB2019}%
  \BibitemOpen
  \bibfield  {author} {\bibinfo {author} {\bibfnamefont {C.~E.}\ \bibnamefont
  {Doss}}, \bibinfo {author} {\bibfnamefont {E.}~\bibnamefont {Adli}}, \bibinfo
  {author} {\bibfnamefont {R.}~\bibnamefont {Ariniello}}, \bibinfo {author}
  {\bibfnamefont {J.}~\bibnamefont {Cary}}, \bibinfo {author} {\bibfnamefont
  {S.}~\bibnamefont {Corde}}, \bibinfo {author} {\bibfnamefont
  {B.}~\bibnamefont {Hidding}},  \emph {et~al.},\ }\href {\doibase
  10.1103/PhysRevAccelBeams.22.111001} {\bibfield  {journal} {\bibinfo
  {journal} {Phys. Rev. Accel. Beams}\ }\textbf {\bibinfo {volume} {22}},\
  \bibinfo {pages} {111001} (\bibinfo {year} {2019})}\BibitemShut {NoStop}%
\bibitem [{\citenamefont {Manwani}\ \emph {et~al.}(2021)\citenamefont
  {Manwani}, \citenamefont {Majernik}, \citenamefont {Yadav}, \citenamefont
  {Hansel},\ and\ \citenamefont {Rosenzweig}}]{manwani2021resonantBunchTrain}%
  \BibitemOpen
  \bibfield  {author} {\bibinfo {author} {\bibfnamefont {P.}~\bibnamefont
  {Manwani}}, \bibinfo {author} {\bibfnamefont {N.}~\bibnamefont {Majernik}},
  \bibinfo {author} {\bibfnamefont {M.}~\bibnamefont {Yadav}}, \bibinfo
  {author} {\bibfnamefont {C.}~\bibnamefont {Hansel}}, \ and\ \bibinfo {author}
  {\bibfnamefont {J.~B.}\ \bibnamefont {Rosenzweig}},\ }\href
  {https://journals.aps.org/prab/abstract/10.1103/PhysRevAccelBeams.24.051302}
  {\bibfield  {journal} {\bibinfo  {journal} {Phys. Rev. Accel. Beams}\
  }\textbf {\bibinfo {volume} {24}},\ \bibinfo {pages} {051302} (\bibinfo
  {year} {2021})}\BibitemShut {NoStop}%
\bibitem [{\citenamefont {Kallos}\ \emph {et~al.}(2006)\citenamefont {Kallos},
  \citenamefont {Muggli}, \citenamefont {Katsouleas}, \citenamefont
  {Yakimenko}, \citenamefont {Stolyarov}, \citenamefont {Pogorelsky},
  \citenamefont {Pavlishin}, \citenamefont {Kusche}, \citenamefont {Babzien},
  \citenamefont {Ben-Zvi} \emph {et~al.}}]{kallos2006resonantBunchTrain}%
  \BibitemOpen
  \bibfield  {author} {\bibinfo {author} {\bibfnamefont {E.}~\bibnamefont
  {Kallos}}, \bibinfo {author} {\bibfnamefont {P.}~\bibnamefont {Muggli}},
  \bibinfo {author} {\bibfnamefont {T.}~\bibnamefont {Katsouleas}}, \bibinfo
  {author} {\bibfnamefont {V.}~\bibnamefont {Yakimenko}}, \bibinfo {author}
  {\bibfnamefont {D.}~\bibnamefont {Stolyarov}}, \bibinfo {author}
  {\bibfnamefont {I.}~\bibnamefont {Pogorelsky}}, \bibinfo {author}
  {\bibfnamefont {I.}~\bibnamefont {Pavlishin}}, \bibinfo {author}
  {\bibfnamefont {K.}~\bibnamefont {Kusche}}, \bibinfo {author} {\bibfnamefont
  {M.}~\bibnamefont {Babzien}}, \bibinfo {author} {\bibfnamefont
  {I.}~\bibnamefont {Ben-Zvi}},  \emph {et~al.},\ }in\ \href
  {https://aip.scitation.org/doi/pdf/10.1063/1.2409178} {\emph {\bibinfo
  {booktitle} {AIP Conference Proceedings}}},\ Vol.\ \bibinfo {volume} {877}\
  (\bibinfo {organization} {American Institute of Physics},\ \bibinfo {year}
  {2006})\ pp.\ \bibinfo {pages} {520--526}\BibitemShut {NoStop}%
\bibitem [{\citenamefont {Ferran~Pousa}\ \emph {et~al.}(2019)\citenamefont
  {Ferran~Pousa}, \citenamefont {Martinez de~la Ossa}, \citenamefont
  {Brinkmann},\ and\ \citenamefont {Assmann}}]{pousa2019compact}%
  \BibitemOpen
  \bibfield  {author} {\bibinfo {author} {\bibfnamefont {A.}~\bibnamefont
  {Ferran~Pousa}}, \bibinfo {author} {\bibfnamefont {A.}~\bibnamefont {Martinez
  de~la Ossa}}, \bibinfo {author} {\bibfnamefont {R.}~\bibnamefont
  {Brinkmann}}, \ and\ \bibinfo {author} {\bibfnamefont {R.~W.}\ \bibnamefont
  {Assmann}},\ }\href
  {https://journals.aps.org/prl/abstract/10.1103/PhysRevLett.123.054801}
  {\bibfield  {journal} {\bibinfo  {journal} {Phys. Rev. Lett.}\ }\textbf
  {\bibinfo {volume} {123}},\ \bibinfo {pages} {054801} (\bibinfo {year}
  {2019})}\BibitemShut {NoStop}%
\bibitem [{\citenamefont
  {Lindstr{\o}m}(2021{\natexlab{b}})}]{lindstrom2021self}%
  \BibitemOpen
  \bibfield  {author} {\bibinfo {author} {\bibfnamefont {C.~A.}\ \bibnamefont
  {Lindstr{\o}m}},\ }\href {https://arxiv.org/abs/2104.14460} {\bibfield
  {journal} {\bibinfo  {journal} {arXiv preprint arXiv:2104.14460}\ } (\bibinfo
  {year} {2021}{\natexlab{b}})}\BibitemShut {NoStop}%
\bibitem [{\citenamefont {Ferran~Pousa}\ \emph {et~al.}(2022)\citenamefont
  {Ferran~Pousa}, \citenamefont {Agapov}, \citenamefont {Antipov},
  \citenamefont {Assmann}, \citenamefont {Brinkmann}, \citenamefont {Jalas}
  \emph {et~al.}}]{pousa2022energy}%
  \BibitemOpen
  \bibfield  {author} {\bibinfo {author} {\bibfnamefont {A.}~\bibnamefont
  {Ferran~Pousa}}, \bibinfo {author} {\bibfnamefont {I.}~\bibnamefont
  {Agapov}}, \bibinfo {author} {\bibfnamefont {S.~A.}\ \bibnamefont {Antipov}},
  \bibinfo {author} {\bibfnamefont {R.~W.}\ \bibnamefont {Assmann}}, \bibinfo
  {author} {\bibfnamefont {R.}~\bibnamefont {Brinkmann}}, \bibinfo {author}
  {\bibfnamefont {S.}~\bibnamefont {Jalas}},  \emph {et~al.},\ }\href
  {https://journals.aps.org/prl/abstract/10.1103/PhysRevLett.129.094801}
  {\bibfield  {journal} {\bibinfo  {journal} {Phys. Rev. Lett.}\ }\textbf
  {\bibinfo {volume} {129}},\ \bibinfo {pages} {094801} (\bibinfo {year}
  {2022})}\BibitemShut {NoStop}%
\bibitem [{\citenamefont {Labat}\ \emph {et~al.}(2022)\citenamefont {Labat},
  \citenamefont {Cadabag}, \citenamefont {Ghaith}, \citenamefont {Irman},
  \citenamefont {Berlioux}, \citenamefont {Berteaud} \emph
  {et~al.}}]{labat2022seeded}%
  \BibitemOpen
  \bibfield  {author} {\bibinfo {author} {\bibfnamefont {M.}~\bibnamefont
  {Labat}}, \bibinfo {author} {\bibfnamefont {J.~C.}\ \bibnamefont {Cadabag}},
  \bibinfo {author} {\bibfnamefont {A.}~\bibnamefont {Ghaith}}, \bibinfo
  {author} {\bibfnamefont {A.}~\bibnamefont {Irman}}, \bibinfo {author}
  {\bibfnamefont {A.}~\bibnamefont {Berlioux}}, \bibinfo {author}
  {\bibfnamefont {P.}~\bibnamefont {Berteaud}},  \emph {et~al.},\ }\href
  {https://doi.org/10.21203/rs.3.rs-1692828/v1} {\bibfield  {journal} {\bibinfo
   {journal} {PREPRINT (Version 1) available at Research Square}\ } (\bibinfo
  {year} {2022})}\BibitemShut {NoStop}%
\bibitem [{\citenamefont {Delbos}\ \emph {et~al.}(2018)\citenamefont {Delbos},
  \citenamefont {Werle}, \citenamefont {Dornmair}, \citenamefont {Eichner},
  \citenamefont {H{\"u}bner}, \citenamefont {Jalas}, \citenamefont {Jolly},
  \citenamefont {Kirchen}, \citenamefont {Leroux}, \citenamefont {Messner}
  \emph {et~al.}}]{delbos2018lux}%
  \BibitemOpen
  \bibfield  {author} {\bibinfo {author} {\bibfnamefont {N.}~\bibnamefont
  {Delbos}}, \bibinfo {author} {\bibfnamefont {C.}~\bibnamefont {Werle}},
  \bibinfo {author} {\bibfnamefont {I.}~\bibnamefont {Dornmair}}, \bibinfo
  {author} {\bibfnamefont {T.}~\bibnamefont {Eichner}}, \bibinfo {author}
  {\bibfnamefont {L.}~\bibnamefont {H{\"u}bner}}, \bibinfo {author}
  {\bibfnamefont {S.}~\bibnamefont {Jalas}}, \bibinfo {author} {\bibfnamefont
  {S.}~\bibnamefont {Jolly}}, \bibinfo {author} {\bibfnamefont
  {M.}~\bibnamefont {Kirchen}}, \bibinfo {author} {\bibfnamefont
  {V.}~\bibnamefont {Leroux}}, \bibinfo {author} {\bibfnamefont
  {P.}~\bibnamefont {Messner}},  \emph {et~al.},\ }\href
  {https://www.sciencedirect.com/science/article/pii/S0168900218301153}
  {\bibfield  {journal} {\bibinfo  {journal} {Nucl. Instrum. Methods. Phys.
  Res. A}\ }\textbf {\bibinfo {volume} {909}},\ \bibinfo {pages} {318}
  (\bibinfo {year} {2018})}\BibitemShut {NoStop}%
\bibitem [{\citenamefont {Cahill}\ \emph {et~al.}(2018)\citenamefont {Cahill},
  \citenamefont {Rosenzweig}, \citenamefont {Dolgashev}, \citenamefont
  {Tantawi},\ and\ \citenamefont {Weathersby}}]{cahill2018high}%
  \BibitemOpen
  \bibfield  {author} {\bibinfo {author} {\bibfnamefont {A.~D.}\ \bibnamefont
  {Cahill}}, \bibinfo {author} {\bibfnamefont {J.~B.}\ \bibnamefont
  {Rosenzweig}}, \bibinfo {author} {\bibfnamefont {V.~A.}\ \bibnamefont
  {Dolgashev}}, \bibinfo {author} {\bibfnamefont {S.~G.}\ \bibnamefont
  {Tantawi}}, \ and\ \bibinfo {author} {\bibfnamefont {S.}~\bibnamefont
  {Weathersby}},\ }\href
  {https://journals.aps.org/prab/abstract/10.1103/PhysRevAccelBeams.21.102002}
  {\bibfield  {journal} {\bibinfo  {journal} {Phys. Rev. Accel. Beams}\
  }\textbf {\bibinfo {volume} {21}},\ \bibinfo {pages} {102002} (\bibinfo
  {year} {2018})}\BibitemShut {NoStop}%
\bibitem [{\citenamefont {Kumada}\ \emph {et~al.}(2003)\citenamefont {Kumada},
  \citenamefont {Iwashita}, \citenamefont {Aoki},\ and\ \citenamefont
  {Sugiyama}}]{kumada2003strongest}%
  \BibitemOpen
  \bibfield  {author} {\bibinfo {author} {\bibfnamefont {M.}~\bibnamefont
  {Kumada}}, \bibinfo {author} {\bibfnamefont {Y.}~\bibnamefont {Iwashita}},
  \bibinfo {author} {\bibfnamefont {M.}~\bibnamefont {Aoki}}, \ and\ \bibinfo
  {author} {\bibfnamefont {E.}~\bibnamefont {Sugiyama}},\ }in\ \href
  {https://ieeexplore.ieee.org/abstract/document/1288751/?casa_token=fMPwnsa3JGsAAAAA:vA2PamIm_1KCF87M0lBzF1-Kj9HHp7gbCoe3MSAiLX3vmF0lAcGE7SpVLqCF2Q6V8eOn6I4A}
  {\emph {\bibinfo {booktitle} {Proceedings of the 2003 particle accelerator
  conference}}},\ Vol.~\bibinfo {volume} {3}\ (\bibinfo {organization} {IEEE},\
  \bibinfo {year} {2003})\ pp.\ \bibinfo {pages} {1993--1995}\BibitemShut
  {NoStop}%
\bibitem [{\citenamefont {Borland}(2000)}]{borland2000elegant}%
  \BibitemOpen
  \bibfield  {author} {\bibinfo {author} {\bibfnamefont {M.}~\bibnamefont
  {Borland}},\ }\href {https://www.osti.gov/biblio/761286} {\emph {\bibinfo
  {title} {Elegant: A flexible SDDS-compliant code for accelerator
  simulation}}},\ \bibinfo {type} {Tech. Rep.}\ (\bibinfo  {institution}
  {Argonne National Lab., IL (US)},\ \bibinfo {year} {2000})\BibitemShut
  {NoStop}%
\bibitem [{\citenamefont {Lehe}\ \emph {et~al.}(2016)\citenamefont {Lehe},
  \citenamefont {Kirchen}, \citenamefont {Andriyash}, \citenamefont {Godfrey},\
  and\ \citenamefont {Vay}}]{Lehe2016FBPIC}%
  \BibitemOpen
  \bibfield  {author} {\bibinfo {author} {\bibfnamefont {R.}~\bibnamefont
  {Lehe}}, \bibinfo {author} {\bibfnamefont {M.}~\bibnamefont {Kirchen}},
  \bibinfo {author} {\bibfnamefont {I.~A.}\ \bibnamefont {Andriyash}}, \bibinfo
  {author} {\bibfnamefont {B.~B.}\ \bibnamefont {Godfrey}}, \ and\ \bibinfo
  {author} {\bibfnamefont {J.~L.}\ \bibnamefont {Vay}},\ }\href {\doibase
  10.1016/j.cpc.2016.02.007} {\bibfield  {journal} {\bibinfo  {journal}
  {Comput. Phys. Commun.}\ }\textbf {\bibinfo {volume} {203}},\ \bibinfo
  {pages} {66} (\bibinfo {year} {2016})}\BibitemShut {NoStop}%
\bibitem [{\citenamefont {Ammosov}\ \emph {et~al.}(1986)\citenamefont
  {Ammosov}, \citenamefont {Delone},\ and\ \citenamefont
  {Krainov}}]{ADK_original}%
  \BibitemOpen
  \bibfield  {author} {\bibinfo {author} {\bibfnamefont {M.~V.}\ \bibnamefont
  {Ammosov}}, \bibinfo {author} {\bibfnamefont {N.~B.}\ \bibnamefont {Delone}},
  \ and\ \bibinfo {author} {\bibfnamefont {V.}~\bibnamefont {Krainov}},\ }\href
  {http://www.jetp.ac.ru/cgi-bin/dn/e_064_06_1191.pdf} {\bibfield  {journal}
  {\bibinfo  {journal} {Sov. Phys. JETP}\ }\textbf {\bibinfo {volume} {64}},\
  \bibinfo {pages} {1191} (\bibinfo {year} {1986})}\BibitemShut {NoStop}%
\bibitem [{\citenamefont {Xu}\ \emph {et~al.}(2016)\citenamefont {Xu},
  \citenamefont {Hua}, \citenamefont {Wu}, \citenamefont {Zhang}, \citenamefont
  {Li}, \citenamefont {Wan} \emph {et~al.}}]{xu2016AdiabaticRamps}%
  \BibitemOpen
  \bibfield  {author} {\bibinfo {author} {\bibfnamefont {X.~L.}\ \bibnamefont
  {Xu}}, \bibinfo {author} {\bibfnamefont {J.~F.}\ \bibnamefont {Hua}},
  \bibinfo {author} {\bibfnamefont {Y.~P.}\ \bibnamefont {Wu}}, \bibinfo
  {author} {\bibfnamefont {C.~J.}\ \bibnamefont {Zhang}}, \bibinfo {author}
  {\bibfnamefont {F.}~\bibnamefont {Li}}, \bibinfo {author} {\bibfnamefont
  {Y.}~\bibnamefont {Wan}},  \emph {et~al.},\ }\href
  {https://journals.aps.org/prl/abstract/10.1103/PhysRevLett.116.124801}
  {\bibfield  {journal} {\bibinfo  {journal} {Phys. Rev. Lett.}\ }\textbf
  {\bibinfo {volume} {116}},\ \bibinfo {pages} {124801} (\bibinfo {year}
  {2016})}\BibitemShut {NoStop}%
\bibitem [{\citenamefont {Knetsch}\ \emph {et~al.}(2021)\citenamefont
  {Knetsch}, \citenamefont {Sheeran}, \citenamefont {Boulton}, \citenamefont
  {Niknejadi}, \citenamefont {P{\~o}der}, \citenamefont {Schaper} \emph
  {et~al.}}]{knetsch2021stable}%
  \BibitemOpen
  \bibfield  {author} {\bibinfo {author} {\bibfnamefont {A.}~\bibnamefont
  {Knetsch}}, \bibinfo {author} {\bibfnamefont {B.}~\bibnamefont {Sheeran}},
  \bibinfo {author} {\bibfnamefont {L.}~\bibnamefont {Boulton}}, \bibinfo
  {author} {\bibfnamefont {P.}~\bibnamefont {Niknejadi}}, \bibinfo {author}
  {\bibfnamefont {K.}~\bibnamefont {P{\~o}der}}, \bibinfo {author}
  {\bibfnamefont {L.}~\bibnamefont {Schaper}},  \emph {et~al.},\ }\href
  {https://journals.aps.org/prab/abstract/10.1103/PhysRevAccelBeams.24.101302}
  {\bibfield  {journal} {\bibinfo  {journal} {Phys. Rev. Accel. Beams}\
  }\textbf {\bibinfo {volume} {24}},\ \bibinfo {pages} {101302} (\bibinfo
  {year} {2021})}\BibitemShut {NoStop}%
\bibitem [{\citenamefont {Telnov}(1995)}]{telnov1995principles}%
  \BibitemOpen
  \bibfield  {author} {\bibinfo {author} {\bibfnamefont {V.}~\bibnamefont
  {Telnov}},\ }\href
  {https://www.sciencedirect.com/science/article/pii/0168900294011737}
  {\bibfield  {journal} {\bibinfo  {journal} {Nucl. Instrum. Methods. Phys.
  Res. A}\ }\textbf {\bibinfo {volume} {355}},\ \bibinfo {pages} {3} (\bibinfo
  {year} {1995})}\BibitemShut {NoStop}%
\bibitem [{\citenamefont {Yakimenko}\ \emph {et~al.}(2019)\citenamefont
  {Yakimenko}, \citenamefont {Meuren}, \citenamefont {Del~Gaudio},
  \citenamefont {Baumann}, \citenamefont {Fedotov}, \citenamefont {Fiuza},
  \citenamefont {Grismayer}, \citenamefont {Hogan}, \citenamefont {Pukhov},
  \citenamefont {Silva} \emph {et~al.}}]{yakimenko2019prospect}%
  \BibitemOpen
  \bibfield  {author} {\bibinfo {author} {\bibfnamefont {V.}~\bibnamefont
  {Yakimenko}}, \bibinfo {author} {\bibfnamefont {S.}~\bibnamefont {Meuren}},
  \bibinfo {author} {\bibfnamefont {F.}~\bibnamefont {Del~Gaudio}}, \bibinfo
  {author} {\bibfnamefont {C.}~\bibnamefont {Baumann}}, \bibinfo {author}
  {\bibfnamefont {A.}~\bibnamefont {Fedotov}}, \bibinfo {author} {\bibfnamefont
  {F.}~\bibnamefont {Fiuza}}, \bibinfo {author} {\bibfnamefont
  {T.}~\bibnamefont {Grismayer}}, \bibinfo {author} {\bibfnamefont {M.~J.}\
  \bibnamefont {Hogan}}, \bibinfo {author} {\bibfnamefont {A.}~\bibnamefont
  {Pukhov}}, \bibinfo {author} {\bibfnamefont {L.~O.}\ \bibnamefont {Silva}},
  \emph {et~al.},\ }\href
  {https://journals.aps.org/prl/abstract/10.1103/PhysRevLett.122.190404}
  {\bibfield  {journal} {\bibinfo  {journal} {Phys. Rev. Lett.}\ }\textbf
  {\bibinfo {volume} {122}},\ \bibinfo {pages} {190404} (\bibinfo {year}
  {2019})}\BibitemShut {NoStop}%
\bibitem [{\citenamefont {Diederichs}\ \emph {et~al.}(2019)\citenamefont
  {Diederichs}, \citenamefont {Mehrling}, \citenamefont {Benedetti},
  \citenamefont {Schroeder}, \citenamefont {Knetsch}, \citenamefont {Esarey},\
  and\ \citenamefont {Osterhoff}}]{diederichs2019positron}%
  \BibitemOpen
  \bibfield  {author} {\bibinfo {author} {\bibfnamefont {S.}~\bibnamefont
  {Diederichs}}, \bibinfo {author} {\bibfnamefont {T.~J.}\ \bibnamefont
  {Mehrling}}, \bibinfo {author} {\bibfnamefont {C.}~\bibnamefont {Benedetti}},
  \bibinfo {author} {\bibfnamefont {C.~B.}\ \bibnamefont {Schroeder}}, \bibinfo
  {author} {\bibfnamefont {A.}~\bibnamefont {Knetsch}}, \bibinfo {author}
  {\bibfnamefont {E.}~\bibnamefont {Esarey}}, \ and\ \bibinfo {author}
  {\bibfnamefont {J.}~\bibnamefont {Osterhoff}},\ }\href
  {https://journals.aps.org/prab/abstract/10.1103/PhysRevAccelBeams.22.081301}
  {\bibfield  {journal} {\bibinfo  {journal} {Phys. Rev. Accel. Beams}\
  }\textbf {\bibinfo {volume} {22}},\ \bibinfo {pages} {081301} (\bibinfo
  {year} {2019})}\BibitemShut {NoStop}%
\end{thebibliography}
%

\end{document}